\documentclass[british]{scrartcl}
\usepackage[T1]{fontenc}
\usepackage{geometry}
\usepackage{amsmath}
\usepackage{graphicx}

\makeatletter

\providecommand{\tabularnewline}{\\}

\makeatother

\usepackage{babel}
\begin{document}

\title{Pair formation of hard core bosons in flat band systems}

\author{Andreas Mielke\thanks{Institut f\"{u}r Theoretische Physik, Universit\"{a}t Heidelberg,
Philosophenweg 19, D-69120 Heidelberg, Germany}}

\date{\today}
\maketitle
\begin{abstract}
Hard core bosons in a large class of one or two dimensional flat band
systems have an upper critical density, below which the ground states
can be described completely. At the critical density, the ground states
are Wigner crystals. If one adds a particle to the system at the critical
density, the ground state and the low lying multi particle states
of the system can be described as a Wigner crystal with an additional
pair of particles. The energy band for the pair is separated from
the rest of the multi-particle spectrum. The proofs use a Gerschgorin
type of argument for block diagonally dominant matrices. In certain
one-dimensional or tree-like structures one can show that the pair
is localised, for example in the chequerboard chain. For this one-dimensional
system with periodic boundary condition the energy band for the pair
is flat, the pair is localised.
\end{abstract}

\section{Introduction}

Strongly correlated bosons on lattices have attracted lots of interest
in the past few years. One of the main reasons is the recent experimental
progress to study such systems in optical lattices, see \cite{greiner2002QuaphatrasuptoMotinsgasultato,bloch2005Ultquagasoptlat,bloch2008Manphyultgas}
and the references therein. Theoretically, interacting bosons on a
lattice are described by the Hubbard model, proposed first to describe
correlated fermions in condensed matter theory \cite{Hubbard63,Kanamori63,Gutzwiller1963}.
Even before it was used in theoretical chemistry to study correlated
$\pi$-electron systems \cite{pariser1953,pople1953}. The bosonic
Hubbard model was to our knowledge first introduced by Fisher et al.
\cite{Fisher1989}. It is expected to show a rich phase diagram including
a Mott insulator and a superfluid phase. 

It is well accepted that in the bosonic Hubbard model repulsively
bound pairs occur \cite{Winkler2006,Petrosyan2007}. They appear for
a sufficiently strong repulsive interaction as dynamically stable
excited states. More recently, pair formation was proposed in the
ground state of the bosonic Hubbard model in some special one-dimensional
lattice structures \cite{Takayoshi2013,Tovmasyan2013,Phillips2014,Pudleiner2015,Gremaud2016,Drescher2017}.
The pair formation occurring here is a collective effect and is caused
by the interplay between the repulsive interaction and the movement
of the particles in these lattice structures. The common feature of
these one-dimensional lattices is that they have a flat band at the
bottom of the single particle spectrum. 

In the present paper we give a rigorous proof for pair formation in
a large class of flat band structures. The class contains as examples
the chequerboard chain \cite{Pudleiner2015,Drescher2017} and its
two-dimensional analogue, the chequerboard lattice. To our knowledge
this is the first proof of pair formation in two-dimensional lattice
structures with flat bands. But the results are more general. The
class contains infinitely many different structures, also including
tree-like or fractal structures; examples for the latter are line
graphs of some Vicsek fractals or Sierpinsky carpets (see e.g. \cite{Bunde94}
and the references therein). The definition of the class uses some
graph theoretical conditions which we introduce later. 

Whereas the bosonic Hubbard model with flat bands has been studied
only recently, the fermionic case has been investigated since 1989,
starting with a pioneering work by Lieb \cite{Lieb89}. Many rigorous
results have been obtained, for a review see \cite{Lieb93a,Tasaki97b,Mielke2015}
and the references therein. Flat band models are of special interest
since in a flat band a very small interaction can yield strong correlation
effects. Independent of the work on the Hubbard model, flat bands
have been studied as well in spin systems, see e.g. \cite{Schulenburg2002,Derzhko2007b}
and the references therein. Standard examples of flat band systems
are the kagom\'{e} lattice or the chequerboard lattice in two dimensions
or similar analogues in one dimension. But the class of models with
flat bands is very large and such lattices can be constructed in any
dimension. Experimentally, it is possible to build flat band systems
using e.g. optical lattices \cite{jo2012} or exciton-polaritons \cite{Masumoto2012}.

For bosons at low temperature, one is interested in flat band systems
where the flat band is the lowest band in the single particle spectrum.
There are two main classes of flat band systems with that property:
Line graphs of bipartite graphs, see e.g. \cite{Mielke1991,Mielke1992a},
which have no gap between the flat band and the rest of the single
particle spectrum, and other decorated lattices, e.g. the ones proposed
by Tasaki \cite{Tasaki92}, which often have a gap between the flat
band and the rest of the spectrum. If one is interested in obtaining
rigorous results, the existence of a gap often simplifies the proofs.
In the present paper we introduce a class of models which interpolate
between these two cases. We investigate line graphs with modified
hoppings. In these models the hopping is reduced on a subset of the
edges. On one subset it is $t>0,$ on the other one it is $t'$ with
$0<t'\leq t$. The models still have a lowest flat band for all $t'$.
A detailed description is given below. The important point here is
that the model contains a tunable parameter $t'$. The lattices have
a gap above the lowest flat band for sufficiently small $t'$ and
no gap for $t'=t$.

Whereas for fermionic systems, many rigorous results are available,
rigorous results for bosons in flat band systems are rare. For line
graphs of two-connected, bipartite plane graphs, Motruk et al. \cite{Motruk2012}
showed that below a critical density the multi particle ground states
of the bosonic Hubbard model with repulsive interaction can be completely
classified. At the critical density, the bosons form a Wigner crystal,
a fact which was already mentioned in \cite{Huber2010} for few special
lattices of this class like the kagom\'{e} lattice. To our knowledge,
there are up to now no general rigorous results above the critical
density for this class of lattices. The aim of the present paper is
to start filling this gap. We investigate what happens if one adds
one additional particle to the system. Further, we take the limit
of a hard core repulsion between the particles, since for weak interaction
pair formation is not expected \cite{Huber2010}. Technically, the
hard core repulsion reduces the Hilbert space dimension and thereby
simplifies the proofs. We show rigorously that in the flat band systems
of our class a pair is formed if one adds one particle. The pair states
form a band that is separated from the rest of the spectrum. In special
one-dimensional lattices or tree-like structures, the pair is localised
and the effective band is flat except at the boundary. Whether this
true for other lattices or in higher dimensions remains open. 

The interesting case, namely the one where $t'=t$ or at least close
to $t$ cannot be reached with the approach used here. We need a small
but finite value for $t'/t$. Nevertheless, we believe that our model
is helpful for a better understanding of the pair formation at larger
values of $t'$, eventually also for $t'=t$. To support that view
we compare our results with other findings for the class of lattices
discussed here, esp. those from \cite{Drescher2017}.

The paper is organised as follows. In the next section we provide,
simply for illustrational purpose, a simple example, which we use
to explain the basic concepts and ideas and the main results. In Sect.
\ref{sec:Definition-of-the} we define the class of graphs we discuss
and the model. We also present some basic results which are valid
at or below the critical density, based on the rigorous work in \cite{Motruk2012}.
In Sect. \ref{sec:Lower-part-of} we discuss the lower part of the
spectrum for a particle number of one particle above the critical
density. The proof is based on a Gerschgorin type of argument. In
Sect. \ref{sec:Eigenstates} we use this result to prove some properties
of the corresponding eigenstates. It is shown that the low lying eigenstates
are linear combinations of localised pair states. A subclass of one-dimensional
or tree-like systems have a local reflection symmetry. For those,
the low lying eigenstates are degenerate except at the boundaries
of the system and contain a localised pair. In the last section, we
discuss possible generalisations of our results.

\section{A simple example: the chequerboard chain}

\begin{figure}
\begin{centering}
\includegraphics[width=0.95\textwidth]{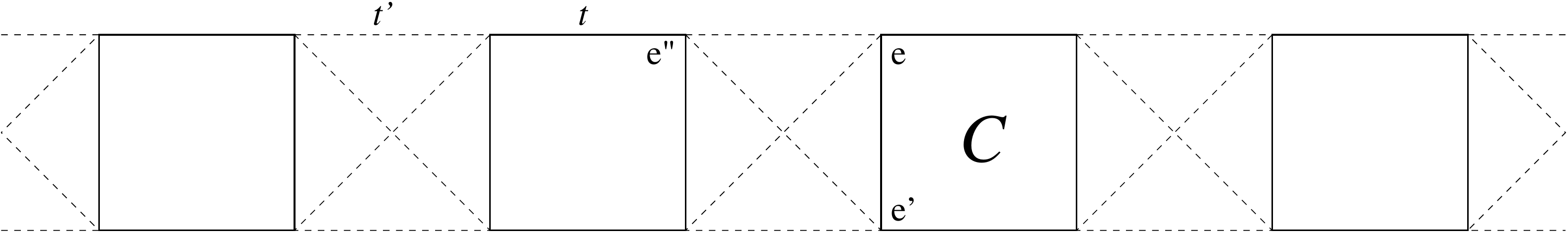}
\par\end{centering}
\caption{\label{fig:cbchain}The chequerboard chain, a chain of cycles $C$
with hoppings $t$ on the cycles and $t'$ between the cycles. }
\end{figure}

The present section explains the basic concepts used in the following
sections within a simple example, the chequerboard chain. The class
of lattices we construct later is quite large, the chequerboard chain
is the most simple example. It is depicted in Fig. \ref{fig:cbchain}.
It consist of cycles, shown with solid lines in the figure, and additional
lines shown as dashed lines connecting neighboured cycles. 

We consider a tight binding model on this lattice with hopping matrix
elements $t$ on the edges belonging to a cycle and $t'$ on the edges
connecting two cycles. We denote the lattice sites by $e$, $e'$.
The Hamiltonian of a single particle moving on this lattice has the
form
\begin{equation}
H_{1}=\sum_{\{e,e'\}}t_{ee'}b_{e}^{\dagger}b_{e'}.\label{eq:HsingleParticle}
\end{equation}
where
\[
t_{ee'}=\begin{cases}
t & \mbox{if }e,\,e'\mbox{ are connected by a line on a cycle.}\\
t' & \mbox{if }e,\,e'\mbox{ are connected by a line between two cycles.}\\
0 & \mbox{otherwise}
\end{cases}
\]
We let $t\geq t'>0$. Consider now a single particle state $\psi_{C}$
which is strictly localised to a single cycle $C$, i.e. $\psi_{C,e}=0$
if the site $e$ does not belong to $C$. For sites $e$ on $C$ the
modulus of $\psi_{C,e}$ shall be the same on all sites and the sign
shall alternate. It is easy to see that this state is an eigenstate
of $H$ with the eigenvalue $-2t$: Take a site $e''$ on a neighbouring
cycle. It is always connected to two sites on $C$. Denote the two
sites $e$ and $e',$ the hopping yields a contribution $t'(\psi_{C,e}+\psi_{C,e'})$
to this site. Since by definition, $\psi_{C,e}$ and $\psi_{C,e'}$
have the same modulus but opposite sign, the contribution vanishes.
Therefore $H\psi_{C}$ vanishes on lattice sites outside $C$. On
$C$, a factor $-2t$ is picked up. Since this construction is possible
for each cycle of the lattice, the eigenvalue $-2t$ of $H_{1}$ is
highly degenerate. The degeneracy is the number of cycles, which is
one quarter of the number of lattice sites. One can show, this will
be done in Sect. \ref{subsec:Basic-properties}, that $-2t$ is the
ground state of $H_{1}$. Since the lattice has four lattice sites
in a unit cell, the energy spectrum consists of four bands. The lowest
band contains the degenerate eigenvalues $-2t,$ it is flat. 

Note that the chequerboard chain has a local reflection symmetry.
A reflection of the cycle $C$ which exchanges the upper and lower
vertices does not change the Hamiltonian. This was already mentioned
and used in \cite{Drescher2017}. We do not use this local reflection
symmetry in our proofs since it is a special property of the chequerboard
chain and some treelike structures we mention later. 

For $t'=0$, the Hamiltonian $H_{1}$ consists of disconnected cycles.
For each cycle, the eigenvalues are $-2t,$ $0$ (twofold degenerate),
and $2t$. Turning on $t'$, the coupling of the cycles lifts the
degeneracy in the upper three bands whereas due to the special structure
of the hopping the lowest band remains flat. As long as $t'<t$, there
is a gap between the lowest band and the other three bands. The gap
closes for $t'=t$. 

Let us now move to the multi particle problem by adding a local repulsive
interaction $U$ to the Hamiltonian. The particles shall be bosons.
The Hamiltonian is
\begin{equation}
H=\sum_{\{e,e'\}}t_{ee'}b_{e}^{\dagger}b_{e'}+U\sum_{e}n_{e}(n_{e}-1).\label{eq:H-1}
\end{equation}

For a sufficiently low density of bosons, we can construct multi particle
ground states simply by putting at most one particle on each cycle
$C$ into the single particle state $\psi_{C}$ of the lowest flat
band. Such a state is clearly a ground state of the hopping part of
the Hamiltonian, since all particles are sitting in single particle
ground states. It minimises also the interaction, because no site
$e$ contains more than one particle. One can show, this was done
in \cite{Motruk2012} for a more general class of lattices, that the
eigenstates constructed that way form a basis of the ground state
space of $H$ when the number of particles is less or equal to the
number of sites. For the present example this is almost trivial. 

The question we want to answer in this paper is what happens if we
add an additional particle to the system so that the number of particles
exceed the number of cycles by one. Since the particles are bosons,
we can put the additional particle in one of the states $\psi_{C}$
and thereby minimise the hopping part of the Hamiltonian. It is clear
that thereby we generate a multi particle state with doubly occupied
sites and the interaction becomes important. We can also spread the
additional particle on all sites. This lowers the interaction energy
but increases the hopping energy. A third possibility is to put a
pair of particles on the same cycle in such a way that they avoid
each other. Then, the interaction energy of the multi particle state
would be zero but the hopping energy would be higher. This is the
favourable option if the interaction energy is large. But it is clear
that due to the presence of $t'$ this is not an exact multi particle
eigenstate of $H$. The question is therefore: Can we characterise
the multi particle ground states of $H$ for large $U$.

Our main result, the proof is given in Sect. \ref{sec:Lower-part-of},
states that for sufficiently small $t'$ the low lying multi particle
states contain indeed a localised pair of bosons. The pair is mainly
localised on a cycle, but not strictly localised. For the chequerboard
chain with periodic boundary conditions, the local reflection symmetry
mentioned above together with the translational invariance guaranties
that the multi particle ground state is degenerate. The degeneracy
is the number of cycles. It can be understood as the original ground
state, in which the added particle forms a pair with one of the other
particles. The pair states are degenerate and can be interpreted as
an effective flat band for the pair. This was already discussed in
\cite{Drescher2017} based on numerical calculations.

For the proof, we use the limit of a hard core repulsion $U\rightarrow\infty$.
Further, for the proof we need a sufficiently small $t'$. But we
expect our result to be true for all $t'\leq t$. This is supported
by the numerical results in \cite{Drescher2017}. 

We now proceed as follows. The next section contains a complete description
of the class of lattices the chequerboard chain is an example of.
We use a graph theoretical language to construct this class. We also
state the basic results for the case where the number of particles
is less or equal to the maximal number which was the number of cycles
in our example. The remaining part of the paper is about the low lying
part of the spectrum and the properties of the corresponding eigenstates
we obtain when we add a particle. The physical picture is exactly
the one proposed above. The additional particle forms a pair with
one of the other particles and the two avoid each other because of
the hard core interaction. The effective pair state is localised,
but not strictly localised.

For the proofs we use a Gerschgorin type argument. The generalised
Gerschgorin theorem we use makes a statement about a matrix with a
block structure where the off-diagonal blocks are sufficiently small.
It can be applied if there is a gap between the ground state and the
rest of the spectrum. If the off-diagonal part is small enough, the
gap remains finite. This is the case if $t'$ is small. 

\section{\label{sec:Definition-of-the}Definition of the model and basic properties}

We consider hard core bosons on a class of graphs which form a subclass
of line graphs of planar graphs. To define the class of lattices we
require some basic notions of graph theory that can be found in the
introductory chapters of the books of Bolobas \cite{Bolobas79} and
Voss \cite{Voss91}. The same construction has been used for fermionic
Hubbard models in \cite{Mielke1991,Mielke1992a}, where more details
are presented. A graph can be drawn as a set of points, called vertices,
and a set of lines, called edges, connecting vertices. A graph is
uniquely defined by the vertex set and the edge set.

\subsection{The class of lattices}

Let $G=(V(G),E(G))$ be a graph with a vertex set $V(G)$ and an edge
set $E(G)$. An edge $e\in E(G)$ is a subset of $V(G)$ with exactly
two elements, the two vertices connected by the edge. We consider
finite graphs, $V(G)$ is a finite set. 

A walk of length $n$ is a sequence $w=(e_{1},e_{2}\ldots,e_{n})$
of edges $e_{i}\in E(G)$ where subsequent edges have exactly one
vertex in common, i.e. $|e_{i}\cap e_{i+1}|=1$ for all $i=1,\ldots,n-1$
and $e_{i}\cap e_{i+1}\cap e_{i+2}=\emptyset$ for all $i=1,\ldots,n-2$.
The first condition also excludes that two subsequent edges are the
same. The second condition assures that the walk passes through the
edge, i.e. the preceding edge connects to one vertex of the edge,
the succeeding edge connects to the second vertex. A path is a self-avoiding
walk which means that no edge is passed more than once by the path,
i.e. $e_{i}\neq e_{j}$ for $i\neq j$. Vertices can be met more than
once by a path. A path which is closed, i.e. $|e_{1}\cap e_{n}|=1$
is called a cycle.

The graph $G$ shall be a planar graph, which means that it can be
drawn in a plane in such a way that no two edges intersect. If, in
the plane representation of $G$, we omit all edges and vertices from
the plane, the plane is decomposed into connected components called
faces. For a finite graph, there is exactly one unbounded face. Let
$F(G)$ be set of bounded faces of the graph. Due to Euler's theorem,
$|F(G)|=|E(G)|-|V(G)|+1$. 

By $C\in F(G)$ we denote a face and also the boundary of that face.
This is clearly possible since each face has a unique boundary. The
boundary $C$ is a cycle. If a cycle is the boundary of a face we
also call it an elementary cycle. Each elementary cycle $C\in F(G)$
itself is a subgraph of $G$ and we denote the vertex set and the
edge set of $C$ by $V(C)$ and $E(C)$ respectively.

Further, we assume that $G$ is two-connected and bipartite. Two-connected
means that the graph remains connected, i.e. does not fall into two
unconnected parts, if an arbitrary edge is removed from $E(G)$. In
other words, each edge belongs to a cycle $C$. Bipartite means that
$V(G)=V_{1}\cup V_{2}$ with $V_{1}\cap V_{2}=\emptyset$ and $|e\cap V_{i}|=1$
for all $e\in E(G)$ and $i=1,2$. If two vertices are connected by
an edge, they are not in the same subset. In a bipartite graph, all
cycles are of even length. 

Let us consider colourings of the faces $F(G)$. Note that the colouring
of the faces of $G$ is equivalent to the vertex colouring of the
dual graph of $G$, see \cite{Bolobas79} for details on colourings.
Two faces $C$ and $C'$ can be coloured with the same colour if they
have no edge in common, $E(C)\cap E(C')=\emptyset$. $\chi(G)$ is
the chromatic number, it is the minimal number of colours needed to
colour the faces of $G$. Since $G$ is planar, $\chi(G)\leq4$, at
most four colours are needed. Let $F_{1}(G)\subset F(G)$ be the largest
set of faces that can be coloured with one colour. If there are several
sets of the same size, $F_{1}(G)$ shall be one of them. Let $E_{1}(G)=\cup_{C\in F_{1}(G)}E(C)\subset E(G)$
be the set of edges contained in the cycles of $F_{1}(G)$. By $G_{1}$
we denote the graph obtained by collecting all the vertices and edges
from all the cycles $C$ in $F(G)$, regarding vertices from different
cycles $C$ as distinct even when they correspond to a single vertex
in $G$. Note that $E(G_{1})=E_{1}(G)$ since each edge in $E_{1}(G)$
belongs to exactly one cycle. Note that $V(G_{1})\neq\cup_{C\in F_{1}(G)}V(C)$
since two cycles in $G$ may contain the same vertex in $V(G)$ but
not in $V(G_{1})$. 

If the faces of $G$ including the unbounded one can be coloured by
two colours, each edge of $G$ belongs to exactly one cycle $C\in F_{1}(G)$
and therefore $E_{1}(G)=E(G)$. Otherwise, $E(G)\backslash E_{1}(G)$
is not empty. The elements of $E(G)\backslash E_{1}(G)$ are called
interstitials. If the faces in $F(G)$ can be coloured with two colours,
interstitials appear only at the boundary of $G$. Otherwise, interstitials
may be everywhere in $G$.

The set of graphs we deal with are line graphs of bipartite planar
graphs. The line graph $L(G)=(V(L(G)),E(L(G)))$ of $G$ is constructed
as follows: $V(L(G))=E(G)$, $E(L(G))=\{\{e,e'\}:\,e,e'\in E(G)\,\mbox{and}\,|e\cap e'|=1\}$.
To draw the line graph, we put a new vertex onto each edge of the
original graph and we connect two new vertices by a new edge, if the
corresponding old edges have a vertex in common. Note that although
the original graph $G$ shall be planar, $L(G)$ is not necessarily
planar. We illustrate the construction below using an example where
$L(G)$ is not planar.

We also need the line graph $L(G_{1})$ of $G_{1}$. Since $G_{1}$
consists of unconnected cycles, $G_{1}$ and $L(G_{1})$ are isomorphic. 

\begin{figure}
\begin{centering}
\begin{tabular}{cccc}
\includegraphics[width=0.22\textwidth]{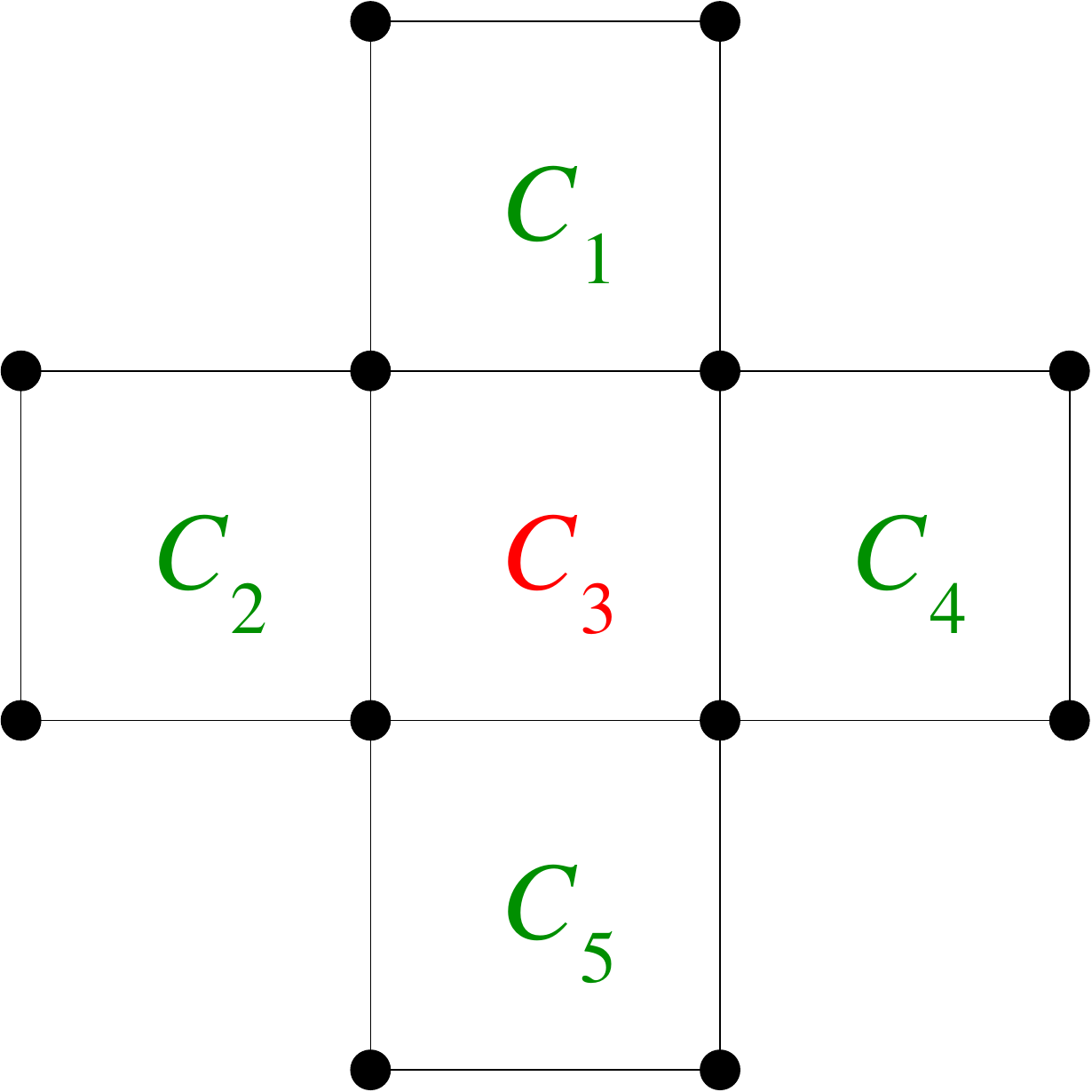} & \includegraphics[width=0.22\textwidth]{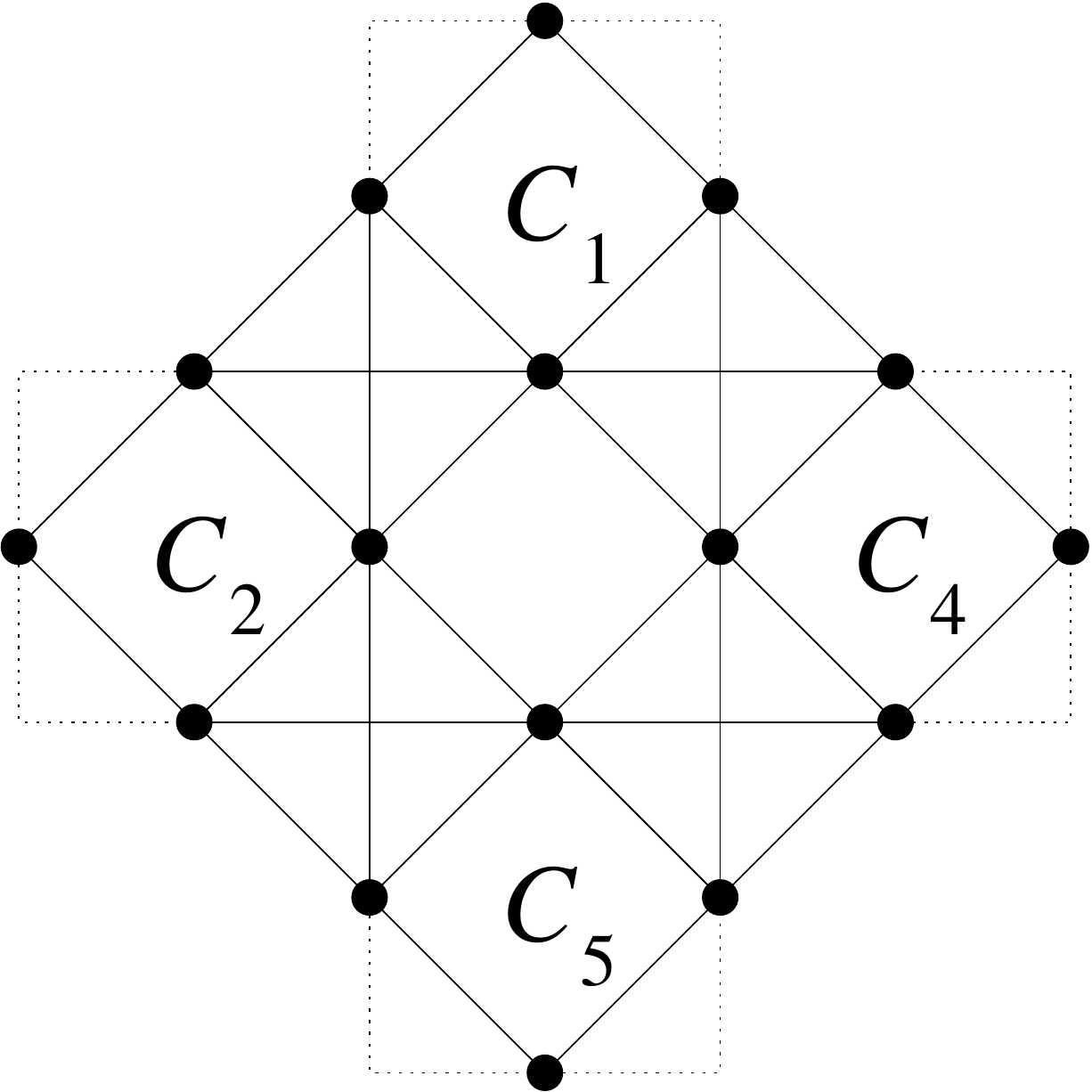} & \includegraphics[width=0.22\textwidth]{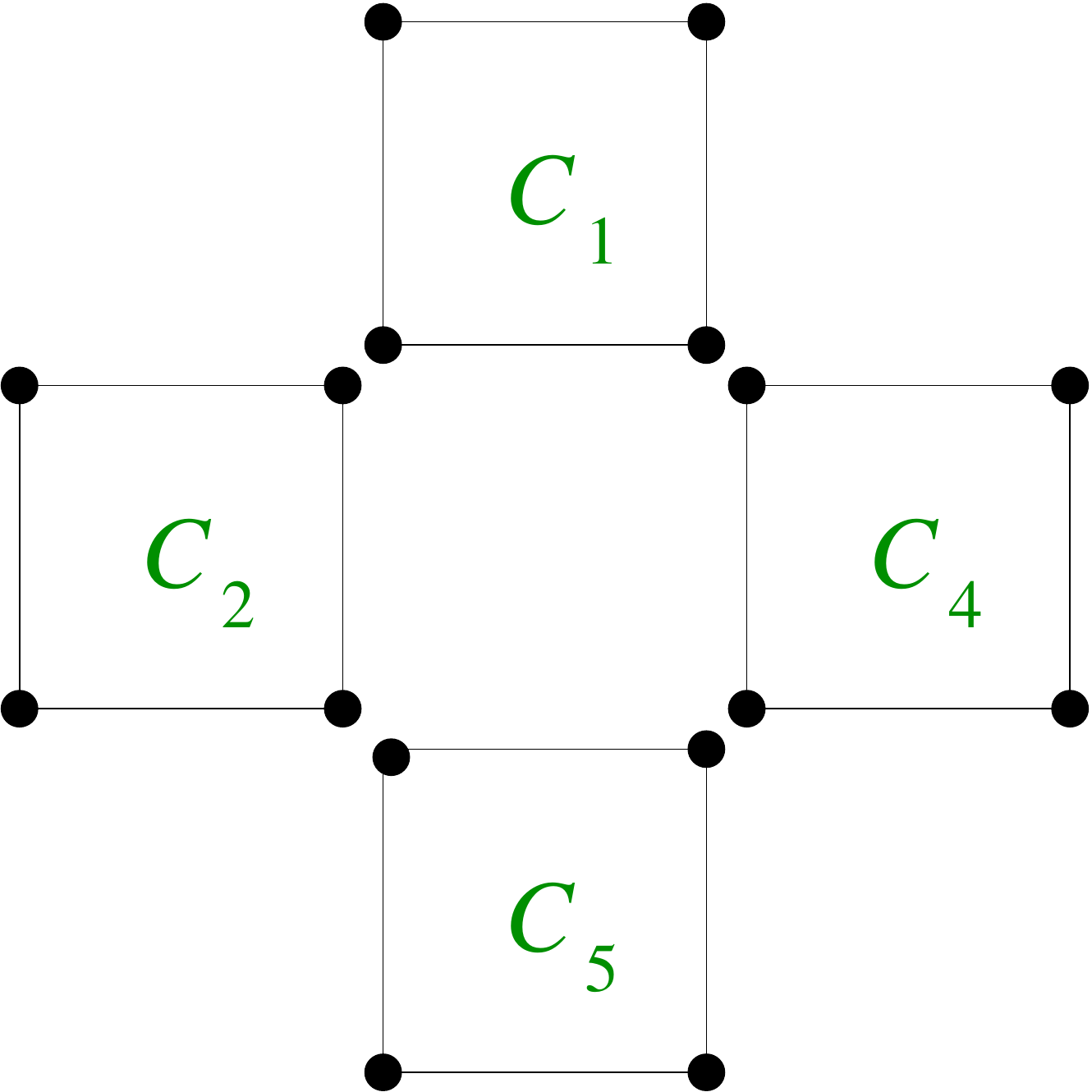} & \includegraphics[width=0.22\textwidth]{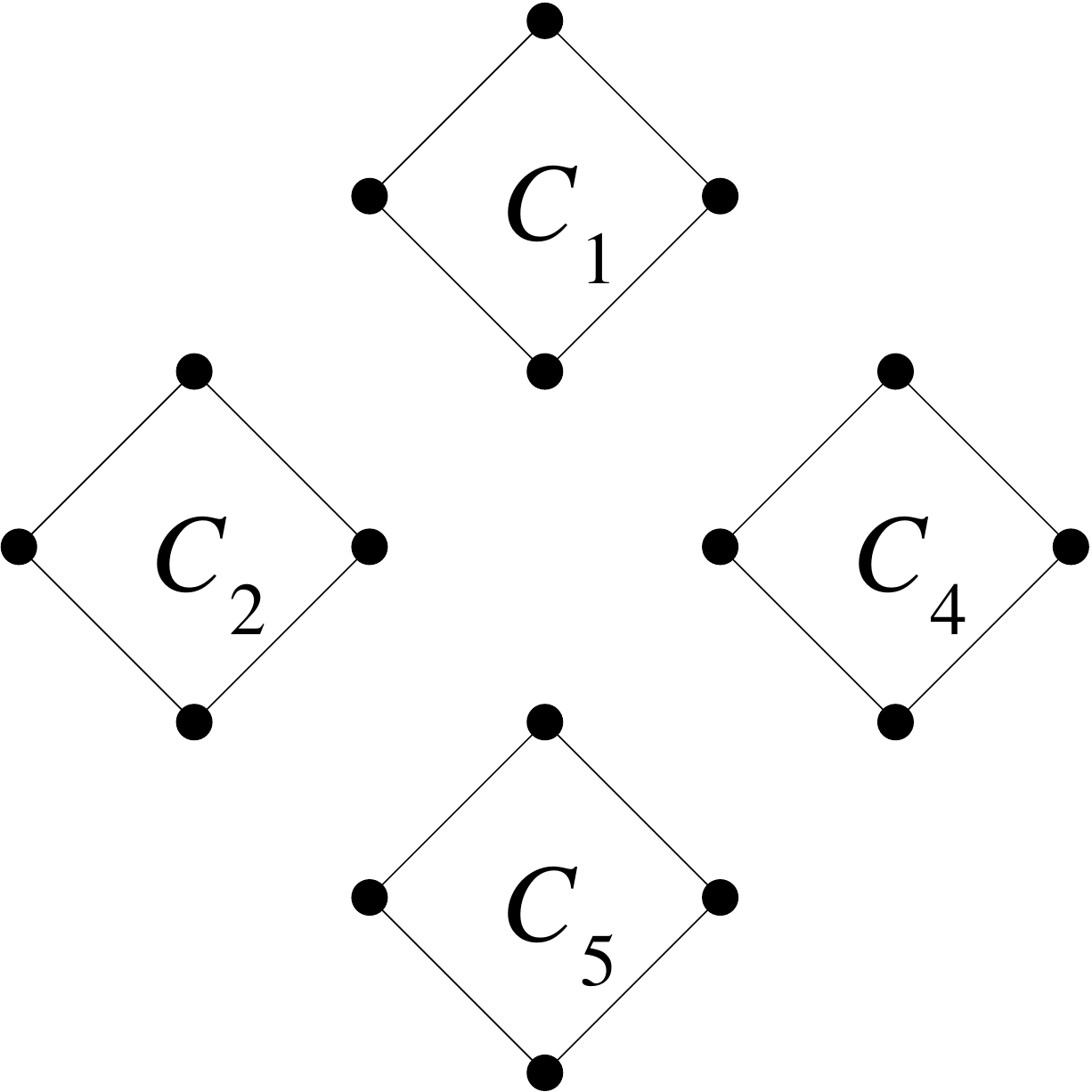}\tabularnewline
\end{tabular}
\par\end{centering}
\caption{\label{fig:lineGraph}From left to right: A graph $G$ and its line
graph $L(G)$ ($G$ with dotted lines). The faces are $C_{i}$. The
graph $G_{1}$ with the four cycles of $F_{1}(G)$ and its line graph
$L(G_{1})$ }
\end{figure}

The construction is illustrated in Fig. \ref{fig:lineGraph}. On the
left the graph $G$ with its set of bounded faces $F(G)=\{C_{1},C_{2},C_{3},C_{4},C_{5}\}$
is shown. Each $C_{i}$ is an elementary cycle. The faces including
the outer face can be coloured with two colours. We have $F_{1}(G)=\{C_{1},C_{2},C_{4},C_{5}\}$.
$L(G_{1})$ on the right hand side consists of the disconnected cycles
$C_{1},\,C_{2},\,C_{4},\,C_{5}$ of $L(G)$. $L(G_{1})$ has the same
vertex set as $L(G)$. There are no interstitials in this example.
It becomes clear that even though $G$ is a planar graph, $L(G)$
in this example is not a planar graph. Whenever $G$ has a vertex
with coordination number larger than 3, the line graph $L(G)$ contains
a complete graph $K_{4}$ as a subgraph, which is not planar. For
our construction we only need that $G$ is planar.

Any bipartite connected planar graph can be used as a starting point.
Therefore, the class of line graphs we are looking at is large. It
contains some well known examples, for instance the kagom\'{e} lattice
and the chequerboard lattice. The latter is the line graph of the
square lattice. For the square lattice, two colours are enough to
colour the bounded faces $F(G)$. Possibly except for some edges at
the boundary, every edge in $E(G)$ belongs to a cycle in $F_{1}(G)$.
The example $G$ depicted in Fig. \ref{fig:lineGraph} is a cutout
of the square lattice. For the honeycomb lattice, three colours are
needed and as a consequence there is a large number of interstitials.
The kagom\'{e} lattice is the line graph of the honeycomb lattice. 

\subsection{The Hamiltonian}

The Hamiltonian of the bosonic Hubbard model on $L(G)$ is defined
as 
\begin{equation}
H=\sum_{\{e,e'\}\in E(L(G))}t_{ee'}b_{e}^{\dagger}b_{e'}+\sum_{e\in V(L(G))}U_{e}n_{e}(n_{e}-1).\label{eq:H}
\end{equation}
We denote the vertices of the line graph $L(G)$ by $e,\,e'$ because
they are the edges of the original graph $G$. We use the usual notation
with creation operators $b_{e}^{\dagger}$ and annihilation operators
$b_{e}$ for bosons on the lattice sites $e\in V(L(G))=E(G)$ with
the usual bosonic commutation relations $[b_{e},b_{e'}]=[b_{e}^{\dagger},b_{e'}^{\dagger}]=0$
and $[b_{e},b_{e'}^{\dagger}]=\delta_{e,e'}$. $n_{e}=b_{e}^{\dagger}b_{e}$
is the particle number on lattice site $e$ and $N=\sum n_{e}$ is
the total particle number, which is conserved. The first part of the
Hamiltonian describes the hopping of the particles along the edges
$\{e,e'\}$ of $L(G)$. We allow only hoppings along the edges, but
the hopping depends on the edge. The second part is the on-site repulsion
$U_{e}>0$. In this paper we let $U_{e}\rightarrow\infty$, which
means that we discuss a model of hard core bosons on $L(G)$. In the
case of a hard core repulsion, at most one particle is allowed on
a site, i.e. $n_{e}\leq1$ for all $e\in E(G)$. Let $P_{\leq1}$
be the projector onto the subspace of states that fulfil this condition.
Then the Hamiltonian can be written as
\begin{equation}
H=P_{\leq1}\sum_{\{e,e'\}\in E(L(G))}t_{ee'}b_{e}^{\dagger}b_{e'}P_{\leq1}.\label{eq:Hprojected}
\end{equation}

\begin{figure}
\begin{centering}
\includegraphics[width=0.3\textwidth]{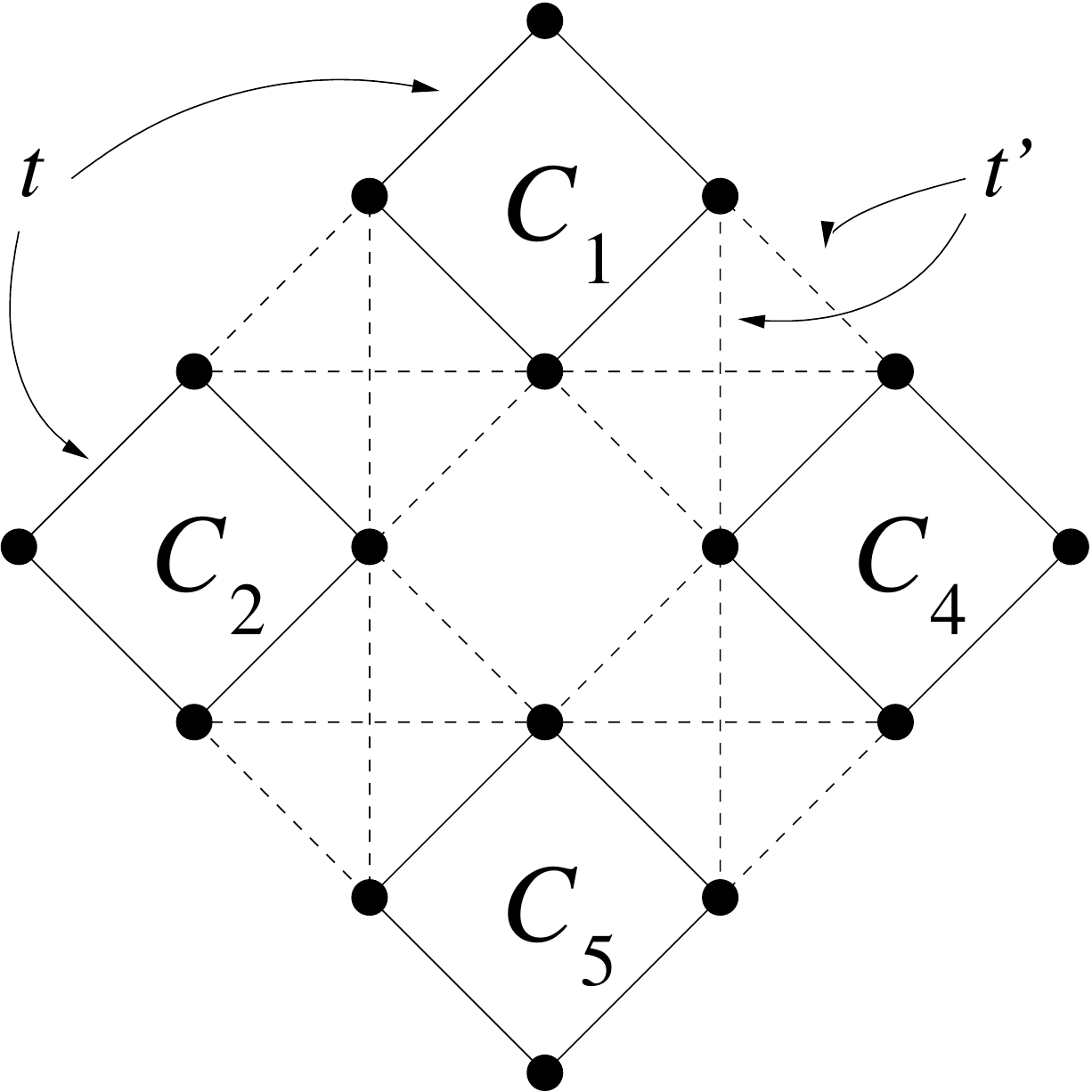}
\par\end{centering}
\caption{\label{fig:The-hoppings-for}The hoppings for the example in Fig.
\ref{fig:lineGraph}}

\end{figure}

The hopping matrix elements are defined as follows:
\begin{equation}
t_{ee'}=\begin{cases}
t & \mbox{if }\{e,e'\}\in E(L(G_{1}))\\
t' & \mbox{if }\{e,e'\}\in E(L(G))\backslash E(L(G_{1}))\\
0 & \mbox{otherwise}
\end{cases}\label{eq:hoppingMatrix}
\end{equation}
and we assume $t\geq t'>0$. Fig. \ref{fig:The-hoppings-for} illustrates
the hoppings for the line graph from Fig. \ref{fig:lineGraph}.

The case $t=t'$ is the usual nearest neighbour hopping on $L(G)$.
The model with $t=t'$ has been treated e.g. in \cite{Motruk2012}.
With this definition of the hopping matrix we write the Hamiltonian
in the form
\begin{equation}
H=tP_{\leq1}\sum_{C\in F_{1}(G)}H_{C}P_{\leq1}+t'P_{\leq1}\sum_{C\neq C'\in F_{1}(G)}H_{C,C'}P_{\leq1}+t'P_{\leq1}H'P_{\leq1}.\label{eq:H=00003DH_C+H_CC'}
\end{equation}
The first part is the Hamiltonian on $L(G_{1})$. $H_{C}$ contains
the hopping on the edges of $C\in F_{1}(G)$ with amplitude 1. $H_{C,C'}$
contains the hoppings from $C'$ to $C$ on the edges connecting the
two cycles, also with amplitude 1. $H'$ is only present if there
are interstitials, it contains all hoppings on paths from some $C$
to some other $C'$ which contain exactly one vertex in $C$, one
vertex in $C'$ and one or more interstitials, again with amplitude
$t'$. 

\subsection{\label{subsec:Basic-properties}Basic properties}

To state some basic properties of the model, we introduce few matrices
often used in graph theory. The most important is the adjacency matrix
$A(G)=(a_{xy})_{x,y\in V(G)}$ where $a_{xy}=1$ if $\{x,y\}\in E(G)$,
$a_{xy}=0$ otherwise. A second important matrix is the vertex-edge
incidence matrix $B(G)=(b_{xe})_{x\in V(G),e\in E(G)}$ where $b_{xe}=1$
if $x\in e$, $b_{xe}=0$ otherwise. The adjacency matrix of $L(G)$
is $A(L(G))=B(G)^{t}B(G)-2$. Since $B(G)^{t}B(G)$ is positive semi-definite,
$A(L(G))$ is bounded from below by $-2.$ The eigenstates of $A(L(G))$
with eigenvalue $-2$ are the elements of the kernel of $B(G)$. They
can be constructed as follows. 

Each bounded face $C\in F(G)$ is bounded by a cycle of even length.
The cycle $C$ can be oriented clockwise. Since $G$ is bipartite,
each edge of $G$ can be oriented to point from one of the two disjoint
subsets of $V(G)$ to the other, e.g. from $V_{1}$ to $V_{2}$. Now
let $v_{C}=(v_{Ce})_{e\in E(G)}$ be defined for a face $C$ of $G$
as follows: $v_{Ce}=1$ if $e\in C$ and $e$ and $C$ have the same
orientation, $v_{Ce}=-1$ if $e\in C$ and $e$ and $C$ have the
opposite orientation, $v_{Ce}=0$ otherwise. It is easy to see that
the $v_{C}$ form a basis of the kernel of $B(G)$, see e.g. \cite{Mielke1991}.
Since by assumption $|F(G)|>0$, the kernel of $B(G)$ is not empty.
We introduce the creation operator $b_{C}^{\dagger}=\frac{1}{\sqrt{|C|}}\sum_{e\in E(C)}v_{Ce}b_{e}^{\dagger}$
and the corresponding annihilation operator $b_{C}=\frac{1}{\sqrt{|C|}}\sum_{e\in E(C)}v_{Ce}b_{e}$. 
\begin{description}
\item [{Proposition.}] For $t\geq t'>0$ the ground state eigenvalue of
the single particle Hamiltonian is $-2t$ and is $|F(G)|$-fold degenerate.
The ground states are $v_{C}$ and their eigenvalues do not depend
on $t'$.
\item [{Proof.}] We have $H+2t=t'(\sum_{C}H_{C}+\sum_{C,C'}H_{C,C'}+H'+2)+(t-t')(\sum_{C}H_{C}+2)$.
The first part is, for a single particle, just $t'(A(L(G))+2)\geq0$.
The second part is non-negative as well and is $(t-t')(A(L(\bigcup_{C\in F_{1}(G)}C))+2)$.
Both are adjacency matrices of line graphs and $F(\bigcup_{C\in F_{1}(G)}C)\subset F(G).$
The states $v_{C}$ minimise both parts and are the only states that
minimise the second part. Thus, by a simple variational argument,
those states are the only single particle ground states of $H$. 
\item [{Remarks.}] The creation operators $b_{C}^{\dagger}$ and the annihilation
operators $b_{C}$ commute with $H_{C,C'}$ and with $H'$. They do
not commute with $P_{\leq1}(\sum_{C,C'}H_{C,C'}+H')P_{\leq1}$.

For $t\geq t'>0$, both parts of $H+2t$ are positive semi-definite.
Therefore, the eigenvalues of $H+2t$ are monotonously increasing
functions of $t'$ and $t-t'$. They are not monotonously increasing
functions of $t'$ for fixed $t$. 

For small values of $t'$ the eigenvalue $-2t$ is separated from
the rest of the single particle spectrum by a gap if $G$ is a sufficiently
large lattice. For $t'=t$, there is no gap.
\item [{Proposition.}] For $t>t'>0$ and $N\leq|F_{1}(G)|$, the ground
states of $H$ are the same as the ones for $t'=0$ and the ground
state energy is $-2tN$. 
\item [{Proof.}] Let $H+2tN=t'P_{\leq1}(\sum_{C}H_{C}+\sum_{C,C'}H_{C,C'}+H'+2N)P_{\leq1}+(t-t')P_{\leq1}(\sum_{C}H_{C}+2N)P_{\leq1}$.
As before, the two parts are positive semi-definite. Again, we use
a simple variational argument. This first part describes hard core
bosons on $L(G)$ with the hopping $t'>0$. For this part, the result
in \cite{Motruk2012} applies. The second part describes hard core
bosons on $L(\bigcup_{C\in F_{1}(G)}C)$. Each of the cycles is disconnected
from the others, the ground states are obtained by putting at most
one particle in a state $v_{C}$. Since $N\leq|F_{1}(G)|$, this is
possible. These states form a subset of the ground states of the first
part and are the only states which minimise the second part. Therefore,
they are the only ground states of $H+2tN$ with eigenvalue 0. Since
they can be obtained by acting with a product of $b_{C}^{\dagger}$
on the vacuum, they do not depend on $t'$.
\item [{Remark.}] We exclude $t'=t$ here because in that case the Hamiltonian
may have more ground states than the Hamiltonian for $t'=0$, since
in that case the second part of $H+2tN$ vanishes and all the states
described in \cite{Motruk2012} become ground states. 
\end{description}

\section{Lower part of the spectrum for $N=|F_{1}(G)|+1$\label{sec:Lower-part-of}}

The results stated in Sect. \ref{subsec:Basic-properties} show that
for $N\leq|F_{1}(G)|$ all ground states of the Hamiltonian can be
constructed. Essentially, the results from \cite{Motruk2012} carry
over to the case $t'\leq t$. For $t'<t$, the set of ground states
is even more simple. The question is what happens for $N>|F_{1}(G)|$.
Clearly, all eigenvalues obey $E\geq-2tN$. 

There is no general answer to that question. The reason is the following.
Consider a graph with exactly two cycles $C_{1}$ and $C_{2}$, both
of length 4, and a long enough chain of interstitials between them.
Let $N=|F(G)|+1=3$. The ground states of $H_{C_{1}}+H_{C_{2}}$ contain
one particle with energy $-2$ on one cycle and two particles on the
other one. The lowest energy of two hard core bosons on a cycle with
length 4 is $-2\sqrt{2}$. Consider now the full Hamiltonian. We can
construct a variational state which contains one particle in each
cycle with energy $-2t$ and one on the chain. For a very long chain,
the lowest energy for a particle on that chain comes close to $-2t'$.
For $t'>(\sqrt{2}-1)t$ it therefore becomes favourable to put the
additional particle on the interstitials. A second point are long
cycles. If a cycle is very long, it is possible to put two particles
on it with only a very small loss of energy. Further, the energies
lie very close to each other. 

Therefore, we first restrict ourselves to a class of graphs without
interstitials and where all cycles $C\in F_{1}(G)$ have length 4.
The chequerboard lattice and the chequerboard chain fall into this
class. Generalisations are discussed in Sect. \ref{sec:Generalisations}. 

On a cycle with length 4 it is easy to construct all eigenstates with
arbitrary particle number between 0 and 4. Since there are only few
different eigenvalues, the eigenstates of $H$ in \eqref{eq:Hprojected}
with $N>|F_{1}(G)|$ are highly degenerate for $t'=0$. For a finite
$t'$, the degeneracy may split and the states will mix. As a consequence,
even for very small $t'>0$ it may be difficult to tell how the spectrum
looks like. In the present case, due to the special structure of the
lattice and the fact that $b_{C}^{\dagger}$ commute with $H_{C,C'},$
it is possible to make use of a variant of Gerschgorin's theorem to
describe the lower part of the spectrum.
\begin{description}
\item [{Theorem~1}] Let $G$ be a connected bipartite planar graph with
$\cup_{C\in F_{1}(G)}E(C)=E(G)$ and $|C|=4$ for all $C\in F_{1}(G)$.
Additionally we assume that all edges of $G$ belong to a cycle $C\in F_{1}(G)$,
. The Hamiltonian $H$ in \eqref{eq:Hprojected} with $N=|F_{1}(G)|+1$
hard core bosons on $L(G)$ has exactly $|F_{1}(G)|$ eigenstates
with an energy at or below $-2t(|F_{1}(G)|-1)-2\sqrt{2}t+\frac{1}{2\sqrt{2}}c(G)t'$
for $t'<\frac{0.14025}{c(G)}t$ where $c(G)$ is the largest number
of cycles in $F_{1}(G)$ connected to some cycle of $F_{1}(G)$. These
eigenvalues are separated from the rest of the spectrum by a finite
gap. 
\end{description}
Note that $c(G)$ is a local quantity. It is not proportional to the
number of lattice sites. For the chequerboard chain in Fig. \ref{fig:cbchain}
and for the example in Fig. \ref{fig:lineGraph} we have $c(G)=2$,
for the chequerboard lattice, $c(G)=4$, independently of the size
of the lattice.

The bound for $t'$ obtained using the Gerschgorin type of argument
is far from being optimal. The reason is that the special structure
of the matrix does not enter. The argument does not take into account
which eigenstates for $t'=0$ can be reached from one of the ground
states at $t'=0$. Since in a perturbational treatment of $t'$ the
first order contribution to the ground state energy vanishes and the
second order yields a negative contribution, we may expect that the
lowest $|F_{1}(G)|$ eigenstates have an energy below $-2t(|F_{1}(G)|-1)-2\sqrt{2}t$.
This is confirmed by the numerical or variational results in \cite{Pudleiner2015,Drescher2017}
for special lattices. Further, the class of graphs is still quite
large. For special graphs in this class, e.g. the chequerboard chain
in one dimension or the chequerboard lattice in two dimensions numerical
results suggest that the result is even true for $t'=t$. 

Note that we only require that the boundaries of the faces in $F_{1}(G)$
have length 4. $G$ may have faces with longer boundaries. Therefore,
line graphs of Vicsek fractals and Sierpinsky carpets \cite{Bunde94}
formed of squares can be treated as well provided they obey the condition
that all edges of $G$ belong to $F_{1}(G)$.

If $t'$ is smaller than the bound given in the theorem there is a
gap between the Gerschgorin cycle around $-2t(|F_{1}(G)|-1)-2\sqrt{2}t$
and the rest of the spectrum. We therefore expect that using the continuity
of the eigenvalues of the Hamiltonian as a function of the parameters
of the Hamiltonian, the fact that there are $|F_{1}(G)|$ low lying
eigenstates separated from the rest of the spectrum remains true for
sufficiently large values of $U$. The main reason is that all quantities
we are dealing with are local.

\subsection{Proof of Theorem 1}

The eigenstates of $P_{\leq1}\sum_{C\in F_{1}(G)}H_{C}P_{\leq1}$
can be constructed from the eigenstates on the cycles of length 4.
For one or three particles, the eigenstates on a single cycle have
the eigenvalues $-2,\,0,\,2$ and 0 is twofold degenerate. For two
particles, the eigenstates on a single cycle have the eigenvalues
$-2\sqrt{2},\,0,\,2\sqrt{2}$, 0 is fourfold degenerate. Four particles
on a single cycle have the eigenvalue 0. The eigenvalues of $P_{\leq1}\sum_{C\in F_{1}(G)}H_{C}P_{\leq1}$
are therefore $E(n_{1},\bar{n}_{1},n_{2},\bar{n}_{2},n_{3},\bar{n}_{3})=-2(n_{1}+n_{3}-\bar{n}_{1}-\bar{n}_{3})-2\sqrt{2}(n_{2}-\bar{n}_{2})$.
$n_{i}$ are the number of cycles with $i$ particles in the lowest
eigenvalue and $\bar{n}_{i}$ are the number of cycles with $i$ particles
in the highest eigenvalue. All other eigenvalues on a single cycle
vanish and therefore do not contribute to the eigenvalues of $P_{\leq1}\sum_{C}H_{C}P_{\leq1}$.
The numbers $n_{i},\,\bar{n}_{i}$ are subject to the additional condition
$n_{1}+\bar{n}_{1}+2n_{2}+2\bar{n}_{2}+3n_{3}+3\bar{n}_{3}\leq N$.
We choose all these states as the basis of the Hilbert space.

We now fix the particle number to $N=|F_{1}(G)|+1$ . The ground states
of $P_{\leq1}\sum_{C}H_{C}P_{\leq1}$ have the eigenvalue $-2(N-2)-2\sqrt{2}$,
there are exactly $|F_{1}(G)|$ of them, corresponding to one doubly
occupied cycle and $N-2$ singly occupied cycles, each in its ground
state. States with higher energies have less singly occupied cycles
in their ground state. The second lowest eigenvalue is $-2(N-4)-4\sqrt{2}$.
The lowest state with $N-n$ singly occupied cycles in their ground
state has the eigenvalue $-2(N-n)-2\sqrt{2}\lfloor\frac{n}{2}\rfloor$.
The second lowest eigenvalue with $N-2$ singly occupied cycles in
their ground state is $-2(N-2)$.

The idea of the proof is to use a Gerschgorin type of argument. We
actually use the generalisation of the Gerschgorin circle theorem
by Feingold and Varga \cite{Feingold1962}. They showed the following.
Let $A$ be the matrix 
\begin{equation}
A=\left(\begin{array}{cccc}
A_{1,1} & A_{1,2} & \ldots & A_{1,N}\\
A_{2,1} & A_{2,2} & \ldots & A_{2,N}\\
\vdots & \vdots & \ddots & \vdots\\
A_{N,1} & A_{N,2} & \ldots & A_{N,N}
\end{array}\right)\label{eq:Ablock}
\end{equation}
where the $A_{i,i}$ are square matrices acting on the subspace $\Omega_{i}$
of order $n_{i}$ and $A_{j,i}$ are $n_{j}\times n_{i}$ matrices.
They show among other things that each eigenvalue $\lambda$ of $A$
satisfies 
\begin{equation}
(||(A_{i,i}-\lambda I_{i})^{-1}||)^{-1}\leq\sum_{k=1,k\neq i}^{N}||A_{i,k}||.\label{eq:Gerschgorin}
\end{equation}
for at least one $i$, $1\leq i\leq N$. Here, $I_{i}$ is the unit
matrix of the same dimension as $A_{ii}$. The matrix norm $||A_{i,j}||$
taken here is derived from an arbitrary vector norm on the subspaces
$\Omega_{i}$ and $\Omega_{j}$ by 
\begin{equation}
||A_{i,j}||=\sup_{x\in\Omega_{j},x\neq0}\frac{||A_{i,j}x||}{||x||}.\label{eq:Matrixnorm}
\end{equation}
One may even choose different norms in the different subspaces $\Omega_{i}$.
We apply this result to our Hamiltonian \eqref{eq:Hprojected}.

\begin{figure}
\centering{}\includegraphics[width=0.3\textwidth]{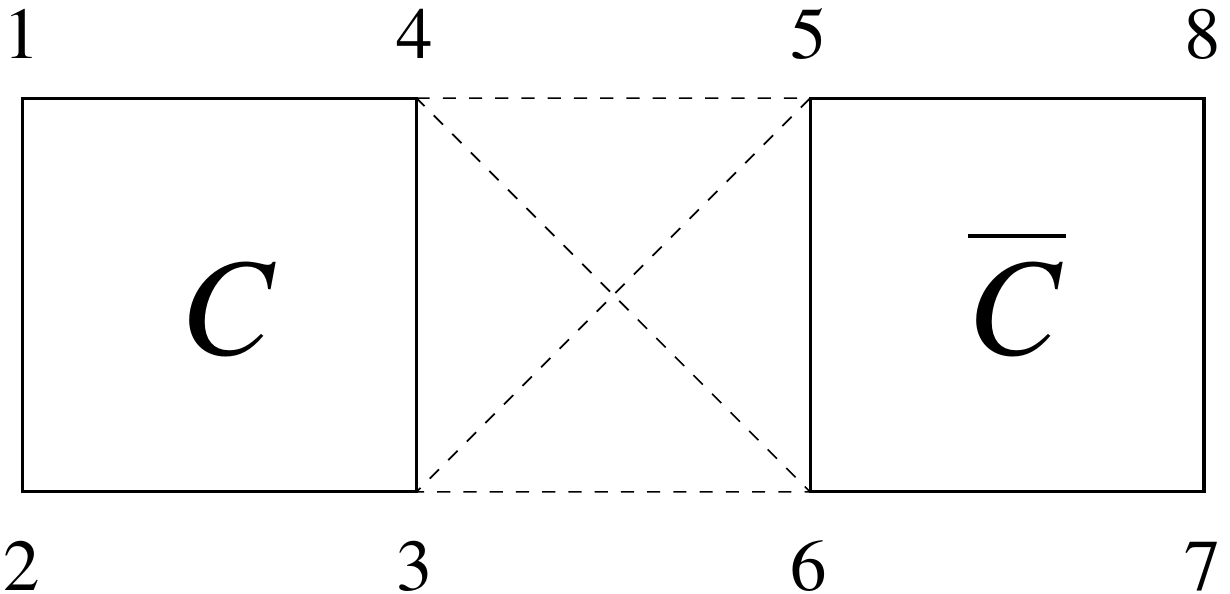}\caption{\label{fig:HCbarC}The cycles $C$ and $\bar{C}$ with the hoppings
contained in $H_{C}$, $H_{\bar{C}}$ and $H_{C,\bar{C}}$. (dashed
lines).}
\end{figure}

Let us first consider the ground states of $P_{\leq1}\sum_{C}H_{C}P_{\leq1}$.
Let $p_{\bar{C}}^{\dagger}$ be the creation operator of the ground
state of two particles on the cycle $\bar{C}$. The ground state of
$P_{\leq1}\sum_{C}H_{C}P_{\leq1}$ with two particles on $\bar{C}$
is
\begin{equation}
\psi_{\bar{C}}=p_{\bar{C}}^{\dagger}\prod_{C\in F_{1}(G)\backslash\{\bar{C}\}}b_{C}^{\dagger}|0\rangle\label{eq:psibarC}
\end{equation}
 We estimate the matrix elements of $P_{\leq1}\sum H_{C,C'}P_{\leq1}$
between this state and other states. 
\begin{align}
P_{\leq1}\sum_{C,C'}H_{C,C'}\psi_{\bar{C}} & =P_{\leq1}\prod_{C''\in F_{1}(G)\backslash\{\bar{C}\}}b_{C''}^{\dagger}\sum_{C}H_{C,\bar{C}}p_{\bar{C}}^{\dagger}|0\rangle\label{eq:mpsibarC}\\
 & =\sum_{C}\prod_{C''\in F_{1}(G)\backslash\{\bar{C},C\}}b_{C''}^{\dagger}P_{\leq1}b_{C}^{\dagger}H_{C,\bar{C}}p_{\bar{C}}^{\dagger}|0\rangle.\nonumber 
\end{align}
The second expression in \eqref{eq:mpsibarC} holds because the cycles
do not overlap. The state $P_{\leq1}b_{C}^{\dagger}H_{C,\bar{C}}p_{\bar{C}}^{\dagger}|0\rangle$
can easily be calculated using the explicit form of the operators
$b_{C}^{\dagger}$, $p_{\bar{C}}^{\dagger}$, and $H_{C,\bar{C}}$,
see Fig. \ref{fig:HCbarC}. In this representation we have 
\begin{equation}
H_{C,\bar{C}}=(b_{3}^{\dagger}+b_{4}^{\dagger})(b_{5}+b_{6}),\label{eq:HCbarC_exp}
\end{equation}
\begin{equation}
p_{\bar{C}}^{\dagger}=\frac{1}{2}(b_{5}^{\dagger}b_{7}^{\dagger}+b_{6}^{\dagger}b_{8}^{\dagger})-\frac{1}{2\sqrt{2}}(b_{5}^{\dagger}b_{6}^{\dagger}+b_{6}^{\dagger}b_{7}^{\dagger}+b_{7}^{\dagger}b_{8}^{\dagger}+b_{8}^{\dagger}b_{5}^{\dagger}),\label{eq:pdaggerbarC}
\end{equation}
and 
\begin{equation}
b_{C}^{\dagger}=\frac{1}{2}(b_{1}^{\dagger}-b_{2}^{\dagger}+b_{3}^{\dagger}-b_{4}^{\dagger}).\label{eq:bdaggerC}
\end{equation}
. 

With the explicit form of these operators we obtain
\begin{equation}
P_{\leq1}[H_{C,\bar{C}},p_{\bar{C}}^{\dagger}]b_{C}^{\dagger}P_{\leq1}=\frac{1}{2}(b_{1}^{\dagger}-b_{2}^{\dagger})(b_{3}^{\dagger}+b_{4}^{\dagger})\left(\frac{1}{2}(1-\frac{1}{\sqrt{2}})(b_{7}^{\dagger}+b_{8}^{\dagger})-\frac{1}{2\sqrt{2}}(b_{5}^{\dagger}+b_{6}^{\dagger})\right)\label{eq:PHpbP}
\end{equation}
and therefore $||P_{\leq1}H_{C,\bar{C}}\psi_{\bar{C}}||_{2}/||\psi_{\bar{C}}||_{2}\leq(1-2^{-1/2})^{1/2}<0.5412$
for the standard norm $||.||_{2}$. But we may take instead the maximum
norm in the basis of the eigenstates of $P_{\leq1}\sum_{C}H_{C}P_{\leq1}$.
The last factor on the right hand side of \eqref{eq:PHpbP} is a sum
of three generators of eigenstates of $H_{\bar{C}}$ with one particle.
The largest prefactor is $\frac{1}{2}$. The other factors yield a
sum of two generators of eigenstates of $H_{C}$ with two particles,
each with prefactor $\frac{1}{\sqrt{2}}$. Therefore we obtain
\begin{equation}
||P_{\leq1}H_{C,\bar{C}}\psi_{\bar{C}}||_{\infty}/||\psi_{\bar{C}}||_{\infty}\leq\frac{1}{2\sqrt{2}},\label{eq:NormHCbarC}
\end{equation}
which yields a slightly better estimate. This holds for every cycle
$C$ connected to $\bar{C}$. Let $c(\bar{C})$ be the number of cycles
connected to the cycle $\bar{C}$. Then we obtain 
\begin{align}
||P_{\leq1}\sum_{C}H_{C,\bar{C}}\psi_{\bar{C}}||_{\infty}/||\psi_{\bar{C}}||_{\infty} & <\sum_{C}||P_{\leq1}H_{C,\bar{C}}\psi_{\bar{C}}||_{\infty}/||\psi_{\bar{C}}||_{\infty}\nonumber \\
 & <\frac{1}{2\sqrt{2}}c(\bar{C}).\label{eq:EstGerschgorin}
\end{align}
Using \eqref{eq:Gerschgorin} this finally yields
\begin{equation}
|-2(N-2)t-2\sqrt{2}t-\lambda|\leq\frac{1}{2\sqrt{2}}c(G)t'\label{eq:firstGerschgorinIntervall}
\end{equation}
where $c(G)=\max_{C\in F_{1}(G)}c(C)$. The centre of all the intervals
is the same, since the eigenvalues of $P_{\leq1}\sum_{C}H_{C}P_{\leq1}$
in theses states are all the same, and all the intervals are contained
in the largest one, which has $c(G)$ on the right hand side.

We now construct subspaces to obtain a suitable block structure of
$H$. For any subset $F\subset F_{1}(G)$ we introduce the subspace
$\Omega_{F}$ which is spanned by the eigenstates of $P_{\leq1}\sum_{C}H_{C}P_{\leq1}$
which are not ground states and which are of the form $\prod_{C\in F}b_{C}^{\dagger}|\psi\rangle$
where $|\psi\rangle$ is any state with $N-|F|$ particles which are
distribute on the remaining cycles in $F_{1}(G)\backslash F$. Further
we introduce the particle number distribution $\bar{n}=(n_{C})_{C\in F_{1}(G)}$
where $n_{C}$ is the number of particle on the face $C$ in the eigenstate
of $P_{\leq1}\sum_{C}H_{C}P_{\leq1}$. Let $\Omega_{F,\bar{n}}\subset\Omega_{F}$
be the subspace with a particle number distribution $\bar{n}.$ Let
$H_{F,\bar{n};F,\bar{n}}$ be the matrix formed by the full Hamiltonian
\eqref{eq:Hprojected} restricted to the subspace $\Omega_{F,\bar{n}}$
and let $H_{F,\bar{n};F'\bar{n}'}$ be the matrix connecting the two
subspaces $\Omega_{F,\bar{n}}$ and $\Omega_{F',\bar{n}'}$. In our
basis, $H_{F,\bar{n};F,\bar{n}}$ is diagonal and the lowest eigenvalue
of $H_{F,\bar{n};F,\bar{n}}$ is $-2t|F|-2\sqrt{2}t\lfloor\frac{1}{2}(N-|F|)\rfloor$,
because we can put $\lfloor\frac{1}{2}(N-|F|)\rfloor$ pairs into
states with the energy $-2\sqrt{2}t$. Let us now look at $H_{C,C'}$
acting on a state out of $\Omega_{F,\bar{n}}$. The important point
is that $H_{C,C'}$ acts only on the cycles $C'$ and $C'\in F_{1}(G)\backslash F$.
We get a non-zero result only if the cycle $C'$ is occupied by some
particles in $|\psi\rangle$. We use a similar representation as in
Fig. \ref{fig:HCbarC} but we allow for an arbitrary state on $C'$.
Further, we have hard-core bosons, the projector eliminates doubly
occupied sites. This yields the rather rough estimate
\begin{equation}
||P_{\leq1}H_{C,C'}\prod_{C''\in F}b_{C''}^{\dagger}\psi||_{2}\leq2||\psi||_{2}\label{eq:normHCC'}
\end{equation}
 for $\psi\in\Omega_{F}$ if $C'\notin F$ and $||H_{C,C'}\psi||=0$
for $\psi\in\Omega_{F}$ if $C'\in F$. 

Since there are at most $N-|F|$ occupied cycles in states $\psi$,
we obtain $||P_{\leq1}\sum H_{C,C'}P_{\leq1}||_{2}\leq2(N-|F|)c(G)$
for states in $\Omega_{F}$. This yields an estimate for the lower
boundary of the Gerschgorin intervals \eqref{eq:Gerschgorin} $-2t|F|-2\sqrt{2}t\lfloor\frac{1}{2}(N-|F|)\rfloor-2t'(N-|F|)c(G)$,
which can be used if $N-|F|>2$. The length of the Gerschgorin intervals
grows, its centre moves to higher energies $\sim(N-|F|)$. For $2t'c(G)>(2-\sqrt{2})t$,
the lower boundary of the intervals moves to lower energies with growing
$N-|F|$. To avoid that we need
\begin{equation}
t'<\frac{2-\sqrt{2}}{2c(G)}t=\frac{0.2928}{c(G)}t.\label{eq:cond_separated}
\end{equation}
The first Gerschgorin interval \eqref{eq:firstGerschgorinIntervall}
is separated from all others if
\begin{equation}
-2(N-2)t-2\sqrt{2}t+\frac{1}{2\sqrt{2}}c(G)t'<-2t|F|-2\sqrt{2}t\lfloor\frac{1}{2}(N-|F|)\rfloor-2t'(N-|F|)c(G).\label{eq:condSeparated}
\end{equation}
 This is fulfilled for all $N-|F|>2$ if 
\begin{equation}
t'<\frac{4-2\sqrt{2}}{(8+\frac{1}{2\sqrt{2}})c(G)}t=\frac{0.14025}{c(G)}t.\label{eq:cond_separated-1}
\end{equation}
holds. The states with $N-|F|=2$ must be treated separately. For
that case, the lowest diagonal element is $-2t|F|$ and we obtain
$t'<0.649t/c(G)$ which is clearly fulfilled if \eqref{eq:cond_separated-1}
holds. 

\section{Eigenstates\label{sec:Eigenstates}}

Let $\Omega_{0}$ be the space spanned by the ground states $\psi_{\bar{C}}$
of $P_{\leq1}\sum_{C}H_{C}P_{\leq1}$ with $N=|F_{1}(G)|+1$ particles
in \eqref{eq:psibarC}. The dimension of $\Omega_{\text{0 }}$is $|F_{1}(G)|$.
Let $P_{0}$ be the projector onto this subspace and let $\bar{P}_{0}=1-P_{0}$
be the projector onto the orthogonal subspace. We write the Hamiltion
$H$ in \eqref{eq:Hprojected} in the form
\begin{equation}
H=\left(\begin{array}{cc}
H_{0} & H_{01}\\
H_{10} & H_{1}
\end{array}\right)=\left(\begin{array}{cc}
P_{0}HP_{0} & P_{0}H\bar{P_{0}}\\
\bar{P_{0}}HP_{0} & \bar{P_{0}}H\bar{P_{0}}
\end{array}\right).\label{eq:Hblock01}
\end{equation}
By construction $H_{0}=[-2(N-2)-2\sqrt{2}]tP_{0}$. The Gerschgorin
interval corresponding to $\Omega_{0}$ calculated with the norm $||.||_{2}$
is $I_{0}=\{\lambda:\,|-2(N-2)t-2\sqrt{2}t-\lambda|\leq(1-2^{-1/2})^{1/2}c(G)t'$
similar to \eqref{eq:firstGerschgorinIntervall}. Further, let $I_{1}$
be the Gerschgorin interval corresponding to $\bar{P}_{0}$. 
\begin{description}
\item [{Theorem~2}] Under the assumptions above and if the two Gerschgorin
intervals do not overlap, i.e. $I_{0}\cap I_{1}=\emptyset$, the eigenstates
$\psi$ of $H$ with eigenvalue $\lambda$ out of $I_{0}$ have the
property $||P_{0}\psi||_{2}>||\bar{P}_{0}\psi||_{2}$. 
\end{description}
To show this, we start with $H_{10}P_{0}\psi+H_{1}\bar{P}_{0}\psi=\lambda\bar{P}_{0}\psi$.
Putting the part acting on $\bar{P}_{0}\psi$ to the left hand side
yields $\bar{P}_{0}\psi=(\lambda\bar{P}_{0}-\bar{P}_{0}H\bar{P}_{0})^{-1}\bar{P}_{0}HP_{0}\psi$.
Taking $||.||_{2}$ on both sides yields 
\begin{equation}
(||(\lambda\bar{P}_{0}-\bar{P}_{0}H\bar{P}_{0})^{-1}||_{2})^{-1}||\bar{P}_{0}\psi||_{2}\leq||\bar{P}_{0}HP_{0}||_{2}||P_{0}\psi||_{2}.\label{eq:separationH0H1}
\end{equation}
Now assume that $||P_{0}\psi||_{2}\leq||\bar{P}_{0}\psi||_{2}$. Then
we would get 
\begin{equation}
(||(\lambda\bar{P}_{0}-\bar{P}_{0}H\bar{P}_{0})^{-1}||_{2})^{-1}\leq||\bar{P}_{0}HP_{0}||_{2}\label{eq:GerschgorinH1}
\end{equation}
which is exactly the condition \eqref{eq:Gerschgorin} for the Gerschgorin
interval corresponding to the subspace given by $\bar{P}_{0}$. Therefore,
$\lambda\in I_{1}$ which contradicts our assumptions $\lambda\in I_{0}$
and $I_{0}\cap I_{1}=\emptyset$. Therefore we must have $||P_{0}\psi||_{2}>||\bar{P}_{0}\psi||_{2}$.
This is the statement in Theorem 2.

The proof works as well if $H$ in \eqref{eq:Hblock01} is split into
more than two blocks.

This theorem means that the eigenstates with eigenvalues in the lowest
Gerschgorin interval are dominated by the ground states of $P_{\leq1}\sum_{C}H_{C}P_{\leq1}$.
The ground states of $P_{\leq1}\sum_{C}H_{C}P_{\leq1}$ are degenerate
and contain localised pairs. Due to the coupling between the cycles
$\propto t'$, two effects occur. First, we can get arbitrary linear
combinations of these states in a low lying eigenstate of the full
Hamiltonian. Second, other states contribute to the low lying eigenstates
as well. But the $|F_{1}(G)|$ low lying eigenstates are dominated
in the above sense by the linear combination of localised pair states,
which may be localised or extended. In this sense, we can speak of
pair formation. The physical interpretation is that the low lying
multi-particle states can be described by a Wigner crystal of $|F_{1}(G)|$
particles in which one particle is replaced by a quasi-particle, the
pair, for which we obtain an effective narrow band that is separated
from the rest of the multi-particle spectrum. The pairs in the eigenstates
are localised in the usual sense if the eigenvalues in the lowest
Gerschgorin interval are degenerate, but this is not a necessary,
only a sufficient condition. This result can in principle be used
to improve the variational states used in \cite{Drescher2017} by
taking linear combinations of the pair states.

An example where the degeneracy and the localisation of the states
in the lowest Gerschgorin interval can be proven is the chequerboard
chain. As discussed in \cite{Drescher2017}, the chequerboard chain
has a local reflection symmetry. Fig. \ref{fig:HCbarC} shows a part
of such a chain. Exchanging the sites 1 and 2 and the sites 3 and
4, the Hamiltonian remains invariant. In the chain, this holds for
each cycle $C\in F_{1}(G)$. The reflection operator $S_{C}$ that
performs this reflection on the cycle $C$ has the eigenvalues $s_{C}=\pm1$.
A singly occupied cycle in the ground state has $s_{C}=-1$, a doubly
occupied cycle in the ground state has $s_{C}=1$. Therefore, the
ground state $\psi_{\bar{C}}$ of $\sum_{C}H_{C}$ with $N=|F_{1}(G)|+1$
particles, two on $\bar{C}$ and one on all the other cycles has a
signature $s_{\bar{C}}=1$, $s_{C}=-1$ for all $C\in F_{1}(G)\backslash\{\bar{C}\}$.
Since the entire Hamiltonian preserves that symmetry, we can restrict
the Hilbert space to all states with that signature. In that Hilbert
space, all the above arguments can be repeated. The only difference
is that the lowest Gerschgorin interval contains only one eigenvalue
that is not degenerate. The corresponding eigenstate, applying Theorem
2, has a localised pair on $\bar{C}$. This holds true for all cycles
$\bar{C}\in F_{1}(G)$, therefore we obtain $|F_{1}(G)|$ eigenstates
with localised pairs. If the chequerboard chain has periodic boundary
conditions, these states are degenerate. For open boundary conditions
we have $c(\bar{C})=1$ for the two cycles at the boundary, $c(\bar{C})=2$
otherwise and therefore we get different eigenvalues for states close
to the boundary. This argument yields a rigorous proof for the statements
in \cite{Drescher2017} for $t'<0.065t$. The numerical results in
\cite{Drescher2017} for $t'=t$ can be repeated for arbitrary $t'<t$
and indicate that the result can be expected to be true for $t'\leq t$.

\begin{figure}
\begin{centering}
\includegraphics[width=0.6\textwidth]{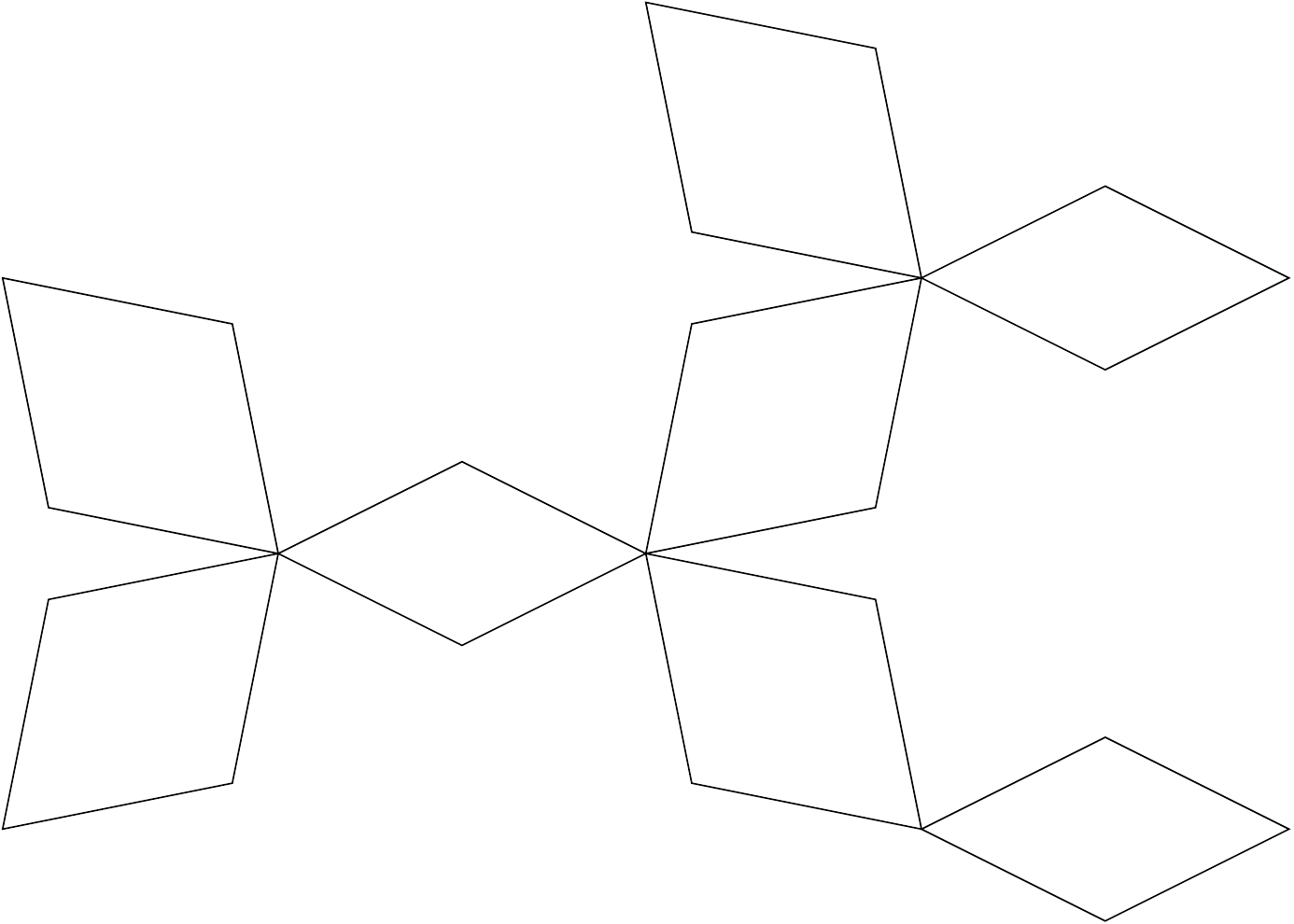}
\par\end{centering}
\caption{\label{fig:treelike}A graph $G$ with a treelike structure and local
reflection symmetry for each elementary cycle.}
\end{figure}

The argument can be readily generalised to tree like structures with
local reflection symmetries. Fig. \ref{fig:treelike} shows an example
of a treelike graph $G$ with local reflection symmetries for each
cycle $C\in F_{1}(G)$. The local symmetry holds for the line graph
$L(G)$ as well. The only differences is that $c(C)>2$ for some cycles
so that $c(G)$ is larger and the value of $t'$ needs to be smaller
in order to have the lowest Gerschgorin interval separated from the
rest of the spectrum. 

For the two dimensional chequerboard lattice, \cite{Pudleiner2015,Drescher2017}
yield arguments on the basis of variational states, extended and localised
ones. The authors show that the localised ones have a lower energy,
which may indicate that in two dimensions localised pairs occur as
well. But we have no rigorous proof for that statement so far.

\section{Generalisations\label{sec:Generalisations}}

There are two possibilities to generalise the above results. The first
is to consider lattices with cycles of length larger than 4 or with
interstitials. The second is to consider $N=|F_{1}(G)|+n$ with some
small $n>1$.

Let us start with larger cycles. For cycles of length 6 one has six
eigenstates with one particle. The two lowest eigenvalues are $-2,\,-1$
and the eigenvalue $-1$ is twofold degenerate. For two particles
on a cycle, the lowest eigenvalue is $-2\sqrt{3}$, the second lowest
is $-\sqrt{3}$. In principle, the technique above is still applicable,
but since the gaps between the states are smaller, the bound for $t'$
becomes smaller as well. 

Unfortunately, this result, although interesting, is not applicable
to the most interesting lattice with cycles of length 6, which is
the kagom\'{e} lattice. For the kagom\'{e} lattice, we have in addition
interstitials, which means that in \eqref{eq:H=00003DH_C+H_CC'} the
third term on the right hand side appears. For the kagom\'{e} lattice,
each interstitial is only connected to two different cycles. This
means that the third term $P_{\leq1}H'P_{\leq1}$ can be decomposed
in the form of the second term $P_{\leq1}\sum_{C.C'\in F_{1}(G)}H_{C,C'}P_{\leq1}$.
This helps a bit, but we would have to take into account that $H_{C,C'}$
contains additional lattice sites and therefore in addition to the
states formed by cycles further states occur. But the important ingredient
of our proof, namely the fact that in the ground states of $P_{\leq1}\sum_{C}H_{C}P_{\leq1}$
only particles from the doubly occupied cycles can hop, remains valid.
Thus we may hope that the proof can be generalised to the kagom\'{e}
lattice. At least the numerical results for the kagom\'{e} chain treated
in \cite{Drescher2017}, which has a local reflection symmetry as
in the case of the chequerboard chain, indicate that there the results
hold true for $t'\leq t$. 

As mentioned before, general graphs with chains of interstitials cannot
be treated and we have good arguments that for those the result is
not valid, see the discussion of Sect. \ref{sec:Lower-part-of}. 

The next question is what happens if we add some more particles. Drescher
et al. \cite{Drescher2017} discussed that question for the chequerboard
chain. Based on their numerical results and based on the exact local
reflection symmetry they argued that for $n=2$ two localised pairs
occur which are well separated from each other. In principle it should
be possible to extend the above method to that case. The lowest Gerschgorin
interval then contains $\frac{1}{2}|F_{1}(G)|(|F_{1}(G)|-1)$ states
corresponding to two doubly occupied cycles and $|F_{1}(G)|-2$ singly
occupied cycles. The estimates are a bit more complicated but still
possible. The upper value for $t'$ to separate the lowest Gerschgorin
cycle from the rest will be lower. We can also use the local reflection
symmetry in this case. This allows to treat a subspace of the entire
Hilbert space in which the lowest Gerschgorin cycle contains only
one eigenvalue. As a consequence, if the lowest Gerschgorin cycle
is separated form the rest of the spectrum, Theorem 2 immediately
shows that the two pairs are localised.

Beside line graphs other flat band systems derived from bipartite
graphs have been proposed \cite{Mielke1991}, which contain tunable
parameters. In these systems, the parameters can be tuned such that
the flat band lies at the bottom of the spectrum and that there is
a gap. These systems are also candidates where the above considerations
can eventually be applied.

\subsection*{Acknowledgement}

I wish to thank Moritz Drescher for fruitful discussion, for reading
the manuscript, and for several helpful remarks.


\begin{thebibliography}{10}
\bibitem{bloch2005Ultquagasoptlat} Bloch, I.: {Ultracold quantum
gases in optical lattices}. \newblock Nature Physics \textbf{1}(1),
23\textendash 30 (2005)

\bibitem{bloch2008Manphyultgas} Bloch, I., Dalibard, J., Zwerger,
W.: {Many-body physics with ultracold gases}. \newblock Reviews
of Modern Physics \textbf{80}(3), 885 (2008)

\bibitem{Bolobas79} Bollob\'{a}s, B.: {Graph theory}. \newblock
Springer Verlag Berlin, Heidelberg, New York 1979

\bibitem{Bunde94}Bunde, A., Havlin, S.: Fractals in Science. Springer
Verlag Berlin, Heidelberg 1994

\bibitem{Derzhko2007b} Derzhko, O., Richter, J., Honecker, A., Schmidt,
H.J.: {Universal properties of highly frustrated quantum magnets
in strong magnetic fields}. \newblock Low. Temp. Phys. \textbf{33},
745 (2007)

\bibitem{Drescher2017} Drescher, M., Mielke, A.: {Hard-core bosons
in flat band systems above the critical density}. \newblock preprint
arXiv:1704.03905 (2017)

\bibitem{Feingold1962} Feingold, D.G., Varga, R.S.: {Block diagonally
dominant matrices and generalizations of the Gerschgorin circle theorem}.
\newblock Pac. J. Math. \textbf{12}(4), 1241\textendash 1250 (1962)

\bibitem{Fisher1989} Fisher, M., Weichman, P., Grinstein, G., Fisher,
D.: {Boson localization and the superfluid-insulator transition}.
\newblock Phys. Rev. B \textbf{40}, 546 (1989)

\bibitem{greiner2002QuaphatrasuptoMotinsgasultato} Greiner, M., Mandel,
O., Esslinger, T., H\"{a}nsch, T.W., Bloch, I.: {Quantum phase transition
from a superfluid to a Mott insulator in a gas of ultracold atoms}.
\newblock nature \textbf{415}(6867), 39\textendash 44 (2002)

\bibitem{Gremaud2016} Gr\'{e}maud, B., Batrouni, G.G.: {Haldane
phase in the sawtooth lattice: Edge states, entanglement and the flat
band}. \newblock Phys. Rev. B \textbf{95}, 165,131 (2017)

\bibitem{Gutzwiller1963} Gutzwiller, M.C.: {Effect of correlation
on the ferromagnetism of transition metals}. \newblock Phys. Rev.
Lett. \textbf{10}(5), 159 (1963)

\bibitem{Hubbard63} Hubbard, J.: {Electron correlations in narrow
energy bands}. \newblock Proc. Roy. Soz. A \textbf{276}, 238 (1963)

\bibitem{Huber2010} Huber, S.D., Altman, E.: {Bose condensation
in flat bands}. \newblock Phys. Rev. B \textbf{82}, 184,502 (2010)

\bibitem{jo2012} Jo, G.B., Guzman, J., Thomas, C.K., Hosur, P., Vishwanath,
A., Stamper-Kurn, D.M.: {Ultracold atoms in a tunable optical kagome
lattice}. \newblock Phys. Rev. Lett. \textbf{108}(4), 45,305 (2012)

\bibitem{Kanamori63} Kanamori, J.: {Electron Correlation and Ferromagnetism
of Transition Metals}. \newblock Prog. Theo. Phys. \textbf{30},
275 (1963)

\bibitem{Lieb89} Lieb, E.H.: {Two Theorems on the Hubbard Model}.
\newblock Phys. Rev. Lett. \textbf{62}, 1201 (1989)

\bibitem{Lieb93a} Lieb, E.H.: {The Hubbard Model: Some Rigorous
Results and Open Problems}. in: \newblock {\em The Hubbard Model.}
Ed: Baeriswyl, D., Campbell, D. K., Carmelo, J. M. P., Guinea, F.,
Louis, E.. NATO ASI Series, pp. 1-19, \newblock {New York: Springer},
1995 

\bibitem{Masumoto2012} Masumoto, N., Kim, N.Y., Byrnes, T., Kusudo,
K., L\"{o}ffler, A., H\"{o}fling, S., Forchel, A., Yamamoto, Y.: {Exciton\textendash polariton
condensates with flat bands in a two-dimensional kagome lattice}.
\newblock New J. of Phys. \textbf{14}, 065,002 (2012)

\bibitem{Mielke1991} Mielke, A.: {Ferromagnetism in the Hubbard
model on line graphs and further considerations}. \newblock J. Phys.
A: Math. Gen. \textbf{24}(14), 3311\textendash 3321 (1991)

\bibitem{Mielke1992a} Mielke, A.: {Exact ground states for the Hubbard
model on the Kagome lattice}. \newblock Journal of Physics A: Mathematical
and General \textbf{25}(16), 4335\textendash 4345 (1992). 

\bibitem{Mielke2015} Mielke, A. {The Hubbard Model and its Properties}.
in: \newblock {\em Many Body Physics: From Kondo to Hubbard.}
Ed: Pavarini, E., Koch, E., and Coleman, P. \newblock {\em Modeling
and Simulation} Vol 5., J\"{u}lich: Forschungszentrum J\"{u}lich
GmbH, 2015

\bibitem{Motruk2012} Motruk, J., Mielke, A.: {Bose-Hubbard model
on two-dimensional line graphs}. \newblock J. Phys. A \textbf{45}(22),
225,206 (2012)

\bibitem{pariser1953} Pariser, R., Parr, R.G.: {A Semi-Empirical
Theory of the Electronic Spectra and Electronic Structure of Complex
Unsaturated Molecules. I.} \newblock J. Chem. Phys. \textbf{21}(3),
466\textendash 471 (1953)

\bibitem{Petrosyan2007} Petrosyan, D., Schmidt, B., Anglin, J.R.,
Fleischhauer, M.: {Quantum liquid of repulsively bound pairs of particles
in a lattice}. \newblock Phys. Rev. A \textbf{76}, 033,606 (2007)

\bibitem{Phillips2014} Phillips, L.G., {De Chiara}, G., \"{O}hberg,
P., Valiente, M.: {Low-energy behaviour of strongly-interacting bosons
on a flat-banded lattice above the critical filling factor}. \newblock
Phys. Rev. B \textbf{91}, 54,103 (2015)

\bibitem{pople1953} Pople, J.A.: {Electron interaction in unsaturated
hydrocarbons}. \newblock Trans. Faraday Soc. \textbf{49}, 1375\textendash 1385
(1953)

\bibitem{Pudleiner2015} Pudleiner, P., Mielke, A.: {Interacting
bosons in two-dimensional flat band systems}. \newblock Eur. Phys.
J. B \textbf{88}, 207 (2015)

\bibitem{Schulenburg2002} Schulenburg, J., Honecker, A., Schnack,
J., Richter, J., Schmidt, H.J.: {Macroscopic magnetization jumps
due to independent magnons in frustrated quantum spin lattices}.
\newblock Phys. Rev. Lett. \textbf{88}, 167,207 (2002)

\bibitem{Takayoshi2013} Takayoshi, S., Katsura, H., Watanabe, N.,
Aoki, H.: {Phase diagram and pair Tomonaga-Luttinger liquid in a
Bose-Hubbard model with flat bands}. \newblock Phys. Rev. A \textbf{88},
063,613 (2013)

\bibitem{Tasaki92} Tasaki, H.: {Ferromagnetism in the Hubbard Models
with Degenerate Single-Electron Ground States}. \newblock Phys.
Rev. Lett. \textbf{69}, 1608 (1992)

\bibitem{Tasaki97b} Tasaki, H.: {From Nagaoka's ferromagnetism to
flat-band ferromagnetism and beyond: An introduction to ferromagnetism
in the Hubbard model}. \newblock cond-mat/9712219 (1997)

\bibitem{Tovmasyan2013} Tovmasyan, M., van Nieuwenburg, E., Huber,
S.: {Geometry induced pair condensation}. \newblock Phys. Rev.
B \textbf{88}, 220,510(R) (2013)

\bibitem{Voss91} Voss, H.J.: {Cycles and Bridges in Graphs}. \newblock
DVW Berlin; Kluwer Dor (1991)

\bibitem{Winkler2006} Winkler, K., Thalhammer, G., Lang, F., Grimm,
R., Denschlag, H.J., Daley, A.J., Kantian, A., B\"{u}chler, H.P.,
Zoller, P.: {Repulsively bound atom pairs in an optical lattice}.
\newblock Nature (London) \textbf{441}, 853 (2006)
\end{thebibliography}
\end{document}